\documentclass[twocolumn,aps,amssymb,prl]{revtex4}
\usepackage{graphicx}
\usepackage{epsfig,amsmath}

\newcommand{\eavg}[1]{\langle{#1}\rangle}

\begin{document}

\title{Local and Non-local Shot Noise in Multiwalled Carbon Nanotubes}

\author{T.~Tsuneta$^{1}$, P.~Virtanen, F.~Wu$^1$,
T.~Wang$^{2}$, T.~T.~Heikkil\"a$^1$, and P.~J.~Hakonen$^1$}
\affiliation{
$^1$LTL,~Helsinki~University~of~Technology, Espoo, Finland \\
$^2$Chinese Academy of Sciences, Beijing, China}

\date{\today} 

\begin{abstract}
We have investigated shot noise in multiterminal, diffusive
multiwalled carbon nanotubes (MWNTs) at 4.2 K over the frequency
$f=600 - 850$ MHz. Quantitative comparison of our data to
semiclassical theory, based on non-equilibrium distribution
functions, indicates that a major part of the noise is caused by a
non-equilibrium state imposed by the contacts. Our data exhibits
non-local shot noise across weakly transmitting contacts while a
low-impedance contact eliminates such noise almost fully. We obtain
$F_{\rm tube}< 0.03$ for the intrinsic Fano factor of our MWNTs.

\end{abstract}
 \bigskip

\maketitle

Multiwalled carbon nanotubes (MWNTs) are miniscule systems, their diameter being only a few nanometers. Yet, in surprisingly many cases their transport properties can be described with incoherent theories, interference effects showing up only through weak localization \cite{Schonenberger,Strunk}. This is in contrast to single-walled tubes, where interference effects dominate, and give rise for example to Fabry-Perot resonances with distinctive features in conductance and current noise \cite{wu07}.

%
When interference effects are washed out, semiclassical analysis
based on non-equilibrium distribution functions is adequate, and the
circuit theory of noise becomes a powerful tool in considering
nanoscale objects \cite{semiclassical,Nazarov02}. This theory makes
it straightforward to calculate current noise of incoherent dots and
wires, and to relate the current noise to the transmission
properties of the corresponding section of the mesoscopic object.
Semiclassical analysis provides a way to make a distinction between
sample and contact effects, and thereby it allows one to investigate
contact phenomena, of which only a little is known in carbon
nanotube systems.

We have investigated the influence of contacts on the shot noise in
multiterminal, diffusive carbon nanotubes. We have made four-lead
measurements on MWNTs in which two middle probes have been employed
for noise measurements. We show that quantitative information can be
obtained from such measurements using semiclassical circuit theory
in the analysis. We find that probes with contact resistance $R_C <
1$ k$\Omega$ act as strongly inelastic probes, resulting in
incoherent, classical addition of noise of two adjacent sections,
while "bad" contacts ($R_C \sim 10 $ k$\Omega$) act as weakly
perturbing probes which need to be analysed on the same footing as
the other parts of the sample. We also find that good contacts
eliminate noise that couples to the probe from a non-neighboring
voltage biased section. In addition, we find from our analysis that
the tubes themselves are quite noise-free, with a Fano-factor
$F_{\rm tube}< 0.03$. As far as we know, our results are the first
shot noise measurements addressing the contact issues in carbon
nanotubes.

\begin{figure}
  \includegraphics{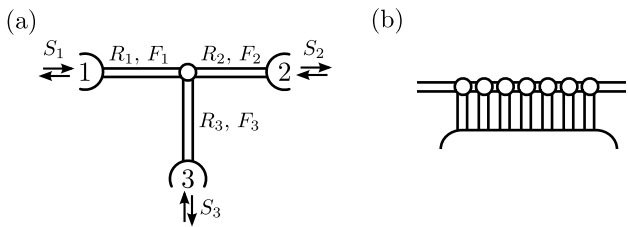}
  \caption{\label{fig:3probe}
    (a) 3-probe structure. The node is denoted by a circle.
    (b) Extended contact.
  }
\end{figure}

To clarify the results of our multi-probe noise measurements, let us
consider the three-terminal structure depicted schematically in
Fig.~\ref{fig:3probe}. Assume that the average current $\eavg{I}$
flows between 1 and 2, and the average potential of the terminal 3
adjusts to the potential of the node. In our work, terminal 3 is
disconnected from the ground at low frequencies, but at the high
frequencies of the noise measurement, the impedance to the ground is
much lower than that of the contacts. As a result, the effect of
voltage fluctuations in the third terminal on the overall noise can
be neglected. We describe two kinds of noise measurements:
``local'', where the noise is measured from one of the terminals 1
or 2, and ``nonlocal'', where the noise is measured from terminal 3.
The shot noise can thus be characterized by the local and non-local
Fano factors, defined as $F_{li}=S_i/e\eavg{I}$, $i=1,2$, and
$F_{nl}=S_3/e\eavg{I}$. Here, $S_i=\int dt\eavg{\delta
  I_i(t)\delta I_i(0)}$ is the low-frequency current noise measured in
terminal $i$.

For strong inelastic scattering inside the node, the resulting
expressions for $F_{l1}$ and $F_{nl}$ would be obtained from the
classical circuit theory, yielding
\begin{align}
  \label{eq:classical3probe}
  F_{l1} = \frac{(G_2+G_3)^2}{G_t^2}F_1
  + \frac{G_1^2}{G_t^2}F_2\,,
  \medspace
  F_{nl} = \frac{G_3^2}{G_t^2}(F_1 + F_2),
\end{align}
where $G_t=G_1+G_2+G_3$. If the nonlocal terminal 3 is well
connected to the node, $G_3\gg G_1, G_2$, the local noise
measurements measure only the local Fano factor, $F_{l1}=F_1$ and
$F_{l2}=F_2$, whereas the nonlocal noise is the sum of them,
$F_{nl}=F_1+F_2$. This is because in this limit the terminal 3
suppresses the voltage fluctuations from the node, and the resulting
noise is only due to the contacts. In the opposite limit
$G_3\rightarrow0$, the nonlocal noise vanishes, and the local noise
is given by the classical addition of voltage fluctuations,
$F_{l1}=F_{l2}=(G_2^2F_1 + G_1^2F_2)/G_t^2$.

At low temperatures the inelastic scattering inside the nanotubes is
suppressed. In this case, assuming that the momentum of the
electrons inside the node is isotropized \cite{BB,Nazarov02}, the
noise can be calculated with the semiclassical Langevin approach
\cite{semiclassical,BB}. It considers the electron energy
distribution function inside each node as a fluctuating quantity
$f(E)=\eavg{f(E)}+\delta{}f(E)$, where $\delta{}f(E)$ are induced by
intrinsic fluctuations of currents between the nodes
$I_{ij}(E)=G_{ij}[f_i(E) - f_j(E)] + \delta I_{ij}(E)$.  Properties
of $\delta I_{ij}$ are known from scattering theory, which allows
calculating all noise correlators in the circuit \cite{BB}.

The fluctuation-averaged energy distribution $f_n(E)$ at the node is
given by a weighted average of (Fermi) distributions at the
terminals, $f_n(E)=\sum_jG_jf_j(E)/G_t$.  In the general case, the
resulting expressions for Fano factors $F_{l1}$, $F_{l2}$, $F_{nl}$
are lengthy, but in the case
of strongly coupled terminal 3, $G_3\gg G_1,G_2$, one obtains the
same expressions as in the classical case.  This is because in this
limit the distribution function of the node is given by the Fermi
function of terminal 3. In the opposite limit $G_3 \ll G_1, G_2$ the
nonlocal noise vanishes and the local noise is given by the
semiclassical sum rule,
\begin{align}
  F_{l1}=F_{l2}
  =\frac{G_2^3 F_1 + G_1^3 F_2 + G_1G_2(G_1+G_2)}{(G_1+G_2)^3} .
  \label{eq:semiclassicalsumrule}
\end{align}
This sum rule applies for any pair of neighboring nodes of an
arbitrary one-dimensional chain of junctions, assuming that
inelastic scattering can be neglected. In the limit of a long chain
with many nodes, applying this rule repeatedly makes the Fano factor
approach the universal value $F=1/3$ characteristic for a diffusive
wire \cite{oberholzer}.

When comparing the semiclassical model to the experiments, we ignore
the resistance of the wires under the contacts and use a model of
localized contacts. In practice, the large width and low interfacial
resistance of the contacts makes the current flow through them
non-uniform. Including this fact in the theoretical model by
describing the tube using a large number of nodes on top of the
contacts (see Fig.~\ref{fig:3probe}b) did not improve the fit
between the model and the experimental noise data, whereas some
improvement was obtained in the fit to the resistance data.
This means that a situation between the localized and
distributed contacts is realized
--- however including this fact in the model would increase
the number of fitting parameters.

 Our individual nanotube samples S1 and S2 were made out of a plasma enhanced CVD
 MWNTs \cite{Koshio} with the length of $L=2.6$ and 5 $\mu$m and the diameter of $\phi=8.9$
nm and 4.0 nm, respectively. The main parameters of the samples are
given in Table \ref{tab:resistances}; the noise data is summarized
in Tables \ref{tab:s1fano} and \ref{tab:s2fano}. The contacts on the
PECVD tubes were made using standard e-beam overlay lithography. In
these contacts, 2 nm of Ti was employed as an adhesive layer before
depositing 30 nm of gold. The width of the four contacts were
$L_{1C} = 400$ nm and $L_{2C} = 550$ nm for samples 1 and 2,
respectively. The strongly doped
Si substrate was employed
as the back gate ($C_g \sim 5$ aF), separated from the sample by
150 nm of SiO$_2$.

\begin{table}
\footnotesize \begin{tabular}{|c|c|c|c|c|c|c|c|c|c|c|c|}
  \hline
    $\phi$ & $L_{12}$ & $L_{23}$ & $L_{34}$ & $R_{12}$ &
    $R_{23}$  & $R_{34}$ & $R_{67}$ & $R_{C1} $ & $R_{C2} $ & $R_{C3}$ & $R_{C4} $ \\
    (nm) & (nm) & (nm) & (nm) & (k$\Omega$) & (k$\Omega$) & (k$\Omega$) &
    (k$\Omega$) & (k$\Omega$) & (k$\Omega$) & (k$\Omega$) & (k$\Omega$)\\
    \hline
  8.9 & 430 & 300 & 540 & 35 & 30 & 34 & 17.5 & - & 0.5 & 12 & - \\
     &  &  &  & 27 & 27 & 41 & 17 & 2.4 & 0.1 & 9.7 & $10^{-5}$ \\
      \hline
  4.0 & 940 & 440 & 1110 & 21 & 25 & 28 & 16.5 & - & 5 & 1.7 & -\\
     & & & & 27& 16 & 31 & 12 & $10^{-5}$ & 2.0 & 1.8 & $10^{-5}$\\
  \hline

\end{tabular}\caption{Main parameters of our samples S1 (upper part) and S2 (lower part).
  The diameter is given by $\phi$ and the length of the tube sections are
  denoted $L_{ij}$ (excluding the range of contacts), where the indices $ij$ correspond to
  pairs of terminals (see Fig.~\ref{exp-setup}) 12, 23, and 34.
  Resistances of the individual sections $R_{ij}$ at $V_g=0$ were measured
  using bias $V=0.1 - 0.2$ V in order to avoid zero bias anomalies;
  $R_{67}$ indicates the 4-wire resistance $R_{4p}$.  Contact resistances $R_{Ck}$,
  weakly dependent on $V_g$ and bias voltage polarity, are given at $V_g=0$ and $V>0$;
  index $k$ identifies the contact. All the resistance data refer to $T= 4.2$ K. The lower row values correspond to
  resistance values obtained in fitting of the semiclassical model. For details, see text.
  }
  \label{tab:resistances}
\end{table}

\begin{figure}

\includegraphics[width=6.5cm]{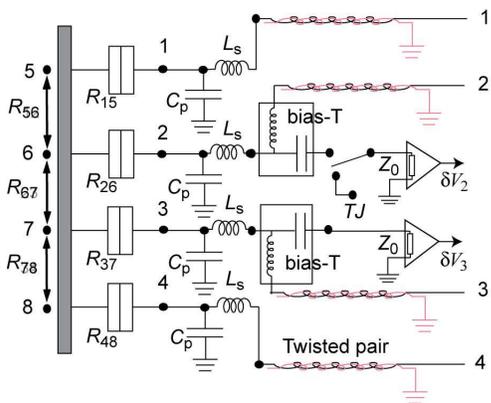}

  \caption{Schematics of our high frequency setup.
    Indices 5-8 refer to nodes with different distribution
    functions on the nanotube. Contacts are drawn as tunnel junctions with
    resistances $R_{ij}$; numbers 1-4 represent the measurement terminals.
    A sum of lead and bonding pad capacitance is given by $C_p \sim 1$ pF while
    the inductors represent bond wires of $L_s \sim 10$ nH. TJ denotes a
    tunnel junction for noise calibration.} \label{exp-setup}
    \end{figure}

Our measurement setup is
illustrated in Fig.~\ref{exp-setup}. Bias-tees are used to separate
dc bias and the bias-dependent noise signal at microwave
frequencies. We use a LHe-cooled, low-noise amplifier
\cite{Cryogenics04} with operating frequency range of $f=600 - 950$
MHz.
A microwave switch and a high-impedance tunnel junction were used to
calibrate the gain.
Our setup measures voltage fluctuations with respect to ground
at the node next to the contact terminal; the voltage fluctuations are converted to
current fluctuations by the contact resistance of the measuring terminal. The sample was
biased using one voltage source, one lead connected to (virtual) ground of the input of
DL1211 current preamplifier and with the remaining two terminals floating.

From four-point measurements at 0.1 - 0.2 V,
we get $R_{4p}=17.5$ k$\Omega$ and $R_{4p}=16.5$  k$\Omega$
(section 6-7 in Fig.~\ref{exp-setup}) for samples S1 and S2,
respectively. Within diffusive transport, this yields for the
resistance per unit length $r_l = 37 - 58$ k$\Omega/\mu$m, which
amounts to $\sim 20$ k$\Omega$ over the length of a contact. The
contact resistances for contacts 2 and 3 of S1 and S2 were
determined as averages from a set of two-lead measurements:
$R_{C2}=(R_{12}+R_{23}-R_{13})/2$ and
$R_{C3}=(R_{23}+R_{34}-R_{24})/2$; this scheme was adopted as there
is a non-local contribution in voltage \cite{Nonlocal}. $R_{C2}$ and
$R_{C3}$ were found to be nearly constant except for a small region
near zero bias. Variation of $V_g=-4 ... +4$ V changes $R_{C_2}$
from $\sim 4$ to 6 k$\Omega$ and $R_{C_3}$ from $\sim 1$ to 3
k$\Omega$ on S2 on average. In both cases, $R_{C}$ increases when
going from $V<0$ to $V>0$ (cf. Table \ref{tab:s2fano}). We cannot
determine the contact resistance of contacts 1 and 4, only the sum
of the contact and the nanotube section. For sample 1, we find
$R_{C2} = 0.1 - 1$ k$\Omega$, indicating that we have an excellent
contact, while $R_{C3} = 12 $ k$\Omega$ points to a weak, tunneling
contact.

\begin{figure}

 \includegraphics[width=6.6cm]{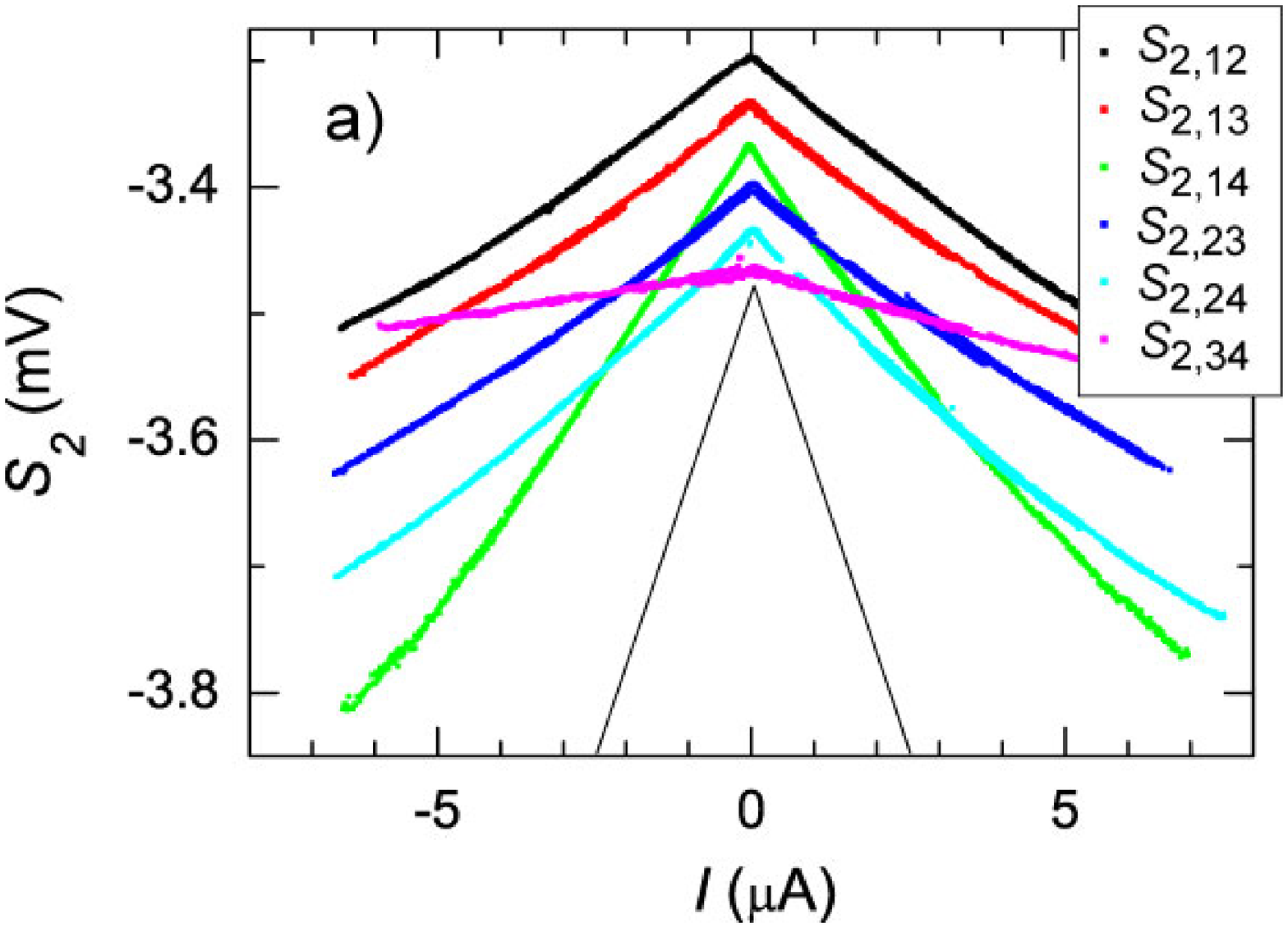}
 \includegraphics[width=6.3cm]{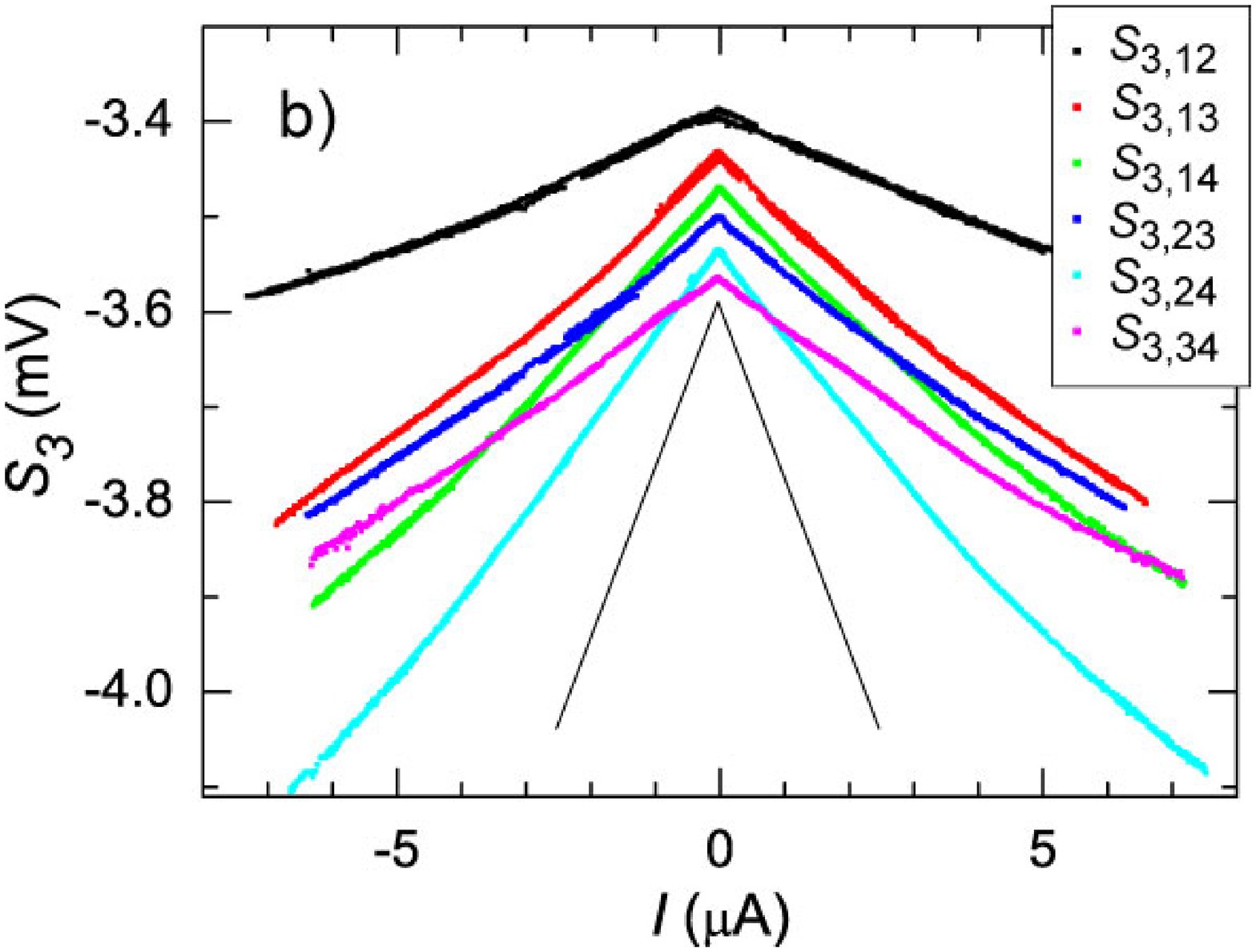}

    \caption{(color online) a) Current noise power (arbitrary units) measured from lead 2
    in sample 2 as a function of bias current, applied according to the label (see text).
    The solid curve indicates noise with $F=1$ (tunnel junction).  b) The
    noise measured from terminal 3 - notations as above.
      } \label{noise-exp3}

\end{figure}

The measured noise as a function of current is displayed in Fig. \ref{noise-exp3} for sample S2. We
measured current noise  $S_{i^2}$ both using DC current, $S_{i^2}= S(I)-S(0)$ and using
AC modulation on top of DC bias: $S_{i^2}=\int_{0}^{I}
    \left(\frac{dS}{dI}\right) dI$, where $S$ represents the
        noise power integrated over the 250 MHz bandwidth
        (divided by 50 $\Omega$) and $\left(\frac{dS}{dI}\right)$ denotes the differentially
measured noise.
As seen in Fig. \ref{noise-exp3}, the measured noise for each
section of the tube is quite well linear with current at $I < 1$
$\mu$A, while at larger currents the Fano factor decreases gradually
with $I$. We determine the Fano factor using linear
fits to $S_{i^2}$ in the range $0.1-2 \mu$A: the results vary over
0.1 - 0.5 as seen in Tables \ref{tab:s1fano} and \ref{tab:s2fano}.

The basic finding of our measurement is that the noise of the sections
adjacent to probes 2 and 3 may behave quite differently, depending
on how strong the contact is between the gold lead and the nanotube.
For terminal 2 of S1, the noise adds in a classical fashion as
expected for a good contact, i.e., $S_{2,13} \simeq
S_{2,12}+S_{2,23}$ ($F_{2,13} \simeq F_{2,12}+F_{2,23}$ in Table
\ref{tab:s1fano}). Here, the first index indicates that noise is
measured from contact 2 while the bias is applied to the terminal
specified by the middle index and the last index tells the grounded
contact.
For the terminal 3 of S1, on the other hand, we find that
$S_{3,24}\simeq S_{3,23} \simeq S_{3,34}$. In sample S2, the results
are intermediate to the above extreme cases (for example, $S_{2,12},
S_{2,23} < S_{2,13}$ and $S_{2,12} + S_{2,23} > S_{2,13}$). All the
above basic relations are in accordance with semiclassical circuit
analysis, while the results related over a good contact can be accounted for by purely
classical circuit theory (cf. Eq. (\ref{eq:classical3probe})).

We fit the semiclassical theory to our data using basically three
types of fitting parameters: the tube resistance per unit length
$r_{l}$, the interfacial resistances $R_{Ck}$ in the contact
regions, and the Fano factor $F_{\rm tube}$ for the parts of the
tube away from contacts. For contacts we use $F=1$ as it makes only
a small contribution to noise in good quality contacts.  The model
thus contains 7 adjustable quantities, and can be used to predict
the measured four resistances and the 12 noise correlators. These
are fit through a least-square minimization procedure.

Tables \ref{tab:s1fano} and \ref{tab:s2fano} display the calculated
results for the noise correlators, while the corresponding
resistances are shown in Table I. In both samples the optimum is
found with $F_{\rm tube} \sim 0$.
The classical model, even though consistent with data with good
contacts, does not provide a good overall account for either of our
samples. The overall agreement of calculated noise with the
measurement is $10 - 30$\%, excluding the two smallest non-local
noise values.
Within the error bars for the measured data, especially due to the difficulties in determining the exact linear-response resistance values (see below), we obtain an upper limit $F_{\rm tube} \lesssim 0.03$, i.e. most of the noise comes from the non-equilibrium state generated by the current (as in the last term in Eq.~\eqref{eq:semiclassicalsumrule}),
not from the transport in the tube itself \cite{NOTE}.

\begin{table}

\begin{tabular}{|c|c|c|c|c|c|c|c|c|}
  \hline
& \multicolumn{4}{|c|}{terminal 2} & \multicolumn{4}{|c|}{terminal 3} \\
\hline
 & $I_{-}$ & $I_{+}$ & Fit & Dev \% & $I_{-}$ & $I_{+}$&  Fit & Dev \%\\
\hline
 \hline
 $S_{k,12}$ & 0.11 & 0.080 & 0.11 & 15 & $<0.005$ & $<0.005$  & 0.006 & 20\\
 $S_{k,13}$ & 0.48 & 0.46 & 0.39 & 16 & 0.41 & 0.39 &  0.38 & 5\\
 $S_{k,14}$ & 0.51 & 0.50 & 0.46 & 9 &  0.50 & 0.46 &  0.45 & 7\\
 $S_{k,23}$ & 0.41 & 0.40 & 0.29 & 28 &  0.42 & 0.38 &  0.38 & 6\\
 $S_{k,24}$ & 0.43 & 0.42 & 0.35 & 17 &  0.54 & 0.47 &  0.45 & 12\\
 $S_{k,34}$ & 0.24 & 0.24 & 0.30 & 26 & 0.46 & 0.50 &  0.40 & 18\\
  \hline
\end{tabular}
\caption{Summary of Fano factors measured at $I<2$ $\mu$A for sample
S1 at terminals $k=2$ and 3. The values in column "Fit" have been
calculated using semiclassical circuit theory (see text), and the
values in column "Dev" show the deviation between the theory and
averaged experimental Fano factor.
}
\label{tab:s1fano}
\end{table}

\begin{table}
\begin{tabular}{|c|c|c|c|c|c|c|c|c|}
  \hline
& \multicolumn{2}{|c|}{$V_{g}=+4V$} & \multicolumn{4}{|c|}{$V_{g}=0$} & \multicolumn{2}{|c|}{$V_{g}=-4V$}\\
\hline
 & $I_{-}$ & $I_{+}$ & $I_{-}$ & $I_{+}$ & Fit & Dev \% & $I_{-}$ & $I_{+}$ \\
\hline
 $R_{C2}$ (k$\Omega$)& 5.8 & 6.4 & 5.3 & 4.9 & 1.8 & 65 & 3.9 & 5.1  \\
 \hline
 $S_{2,12}$ & 0.26 & 0.24 & 0.21 & 0.20 & 0.19 & 10 & 0.24 & 0.23 \\
 $S_{2,13}$ & 0.37 & 0.34 & 0.33 & 0.32 & 0.35 & 6 & 0.38 & 0.35 \\
 $S_{2,14}$ & 0.42 & 0.38 & 0.37 & 0.36 & 0.41 & 11 & 0.36 & 0.36 \\
 $S_{2,23}$ & 0.25 & 0.23 & 0.24 & 0.25 & 0.23 & 5 & 0.25 & 0.24 \\
 $S_{2,24}$ & 0.36 & 0.32 & 0.27 & 0.28 & 0.29 & 7 & 0.28 & 0.21 \\
 $S_{2,34}$ & 0.075 & 0.11 & 0.061 & 0.082 & 0.12 & 64 & 0.055 & 0.053 \\
  \hline
\end{tabular}

\begin{tabular}{|c|c|c|c|c|c|c|c|c|}
  \hline
 & \multicolumn{2}{|c|}{$V_{g}=+4V$} & \multicolumn{4}{|c|}{$V_{g}=0$} & \multicolumn{2}{|c|}{$V_{g}=-4V$}\\
\hline
 & $I_{-}$ & $I_{+}$ & $I_{-}$ & $I_{+}$ & Fit & Dev \% & $I_{-}$ & $I_{+}$ \\
\hline
 $R_{C3}$ (k$\Omega$)& 2.4 & 3.3 & 1.4 & 2.1 & 1.7 & $<$5 & 0.9 & 2.9 \\
 \hline
 $S_{3,12}$ & 0.081 & 0.079 & 0.13 & 0.13 & 0.13 & 2 & 0.043 & 0.052 \\
 $S_{3,13}$ & 0.14 & 0.15 & 0.29 & 0.27 & 0.29 & 4 & 0.14 & 0.14 \\
 $S_{3,14}$ & 0.34 & 0.24 & 0.31 & 0.30 & 0.40 & 30 & 0.20 & 0.21 \\
 $S_{3,23}$ & - & - & 0.24 & 0.25 & 0.23 & 8 & - & - \\
 $S_{3,24}$ & 0.26 & 0.19 & 0.40 & 0.35 & 0.33 & 12 & 0.17 & 0.18\\
 $S_{3,34}$ & 0.12 & 0.087 & 0.24 & 0.19 & 0.17 & 23 & 0.085 & 0.077 \\

  \hline
\end{tabular}
\caption{Summary of Fano factors measured at $I<2$ $\mu$A for S2 at three
different gate voltage values as well as the measured contact
resistances for sample 2 at terminals 2 (top) and 3 (bottom).
Columns "Fit" and "Dev" refer to theoretical fit using semiclassical
analysis and its deviation from the experimental data as in Table
\ref{tab:s1fano}.} \label{tab:s2fano}
\end{table}

We get from Table \ref{tab:s2fano} average Fano factors $F=0.26$ and
$F= 0.18$ at $V_g=-4, 0, +4$ V for terminals 2 and 3, respectively.
These values correspond to ensemble averaged values which describe
the combined properties of the tube and its contacts. On the other
hand, if we take only the noise from single sections 12 and 23 (or
23 and 34), then the average $F=0.24$ (0.16), pretty close to the
above values. Altogether, the variation of local, single section
measurements is 0.10-0.48, quite distinct from simple diffusive wire
expectations, and the semiclassical analysis is able to account for
all these under the premise that the tube is noise-free! This
conclusion is in line with results in SWNT bundles that have shown
small noise as well \cite{Roche02}. In semiconducting MWNTs rather
large noise at 1 GHz has been observed, which has been assigned to
the presence of Schottky barriers \cite{WuSemi}.

The fitted resistances are slightly off from the measured values.
This must be partly because in nanotubes it is difficult to avoid
uncertainties in four-probe measurements (current goes in part
through the voltage probes), and partly because the presence of
non-local voltages. In addition, the quality of our MWNTs may be so
good that the conduction becomes semiballistic and our analysis is
not valid any more.
For example,
in section 1-2 of S1, the IV curve displays a power law which
clearly differs from the rest of the sample.
The fitted contact resistances range over $R_C=0.1-10$ k$\Omega$,
which is in agreement with typical measured values
\cite{Schonenberger,deHeer}.

In summary, we have investigated experimentally shot noise of
multiterminal MWNTs under several biasing conditions. The noise was
found to depend strongly on the contact resistance. At small
interfacial resistance, our 0.4-0.5 micron contacts acted as
inelastic probes and classical noise analysis was found sufficient.
Weaker contacts could be accounted for by using semiclassical
theory. The latter allows to comprehend the observed, broad spectrum
of Fano-factors, but it leads to the conclusion that the intrinsic
noise of MWNTs is nearly zero, at most $F_{\rm tube}< 0.03$. Most of
the observed noise is generated by metal-nanotube contacts, which
govern the non-equilibrium distributions of charge carriers on the
tube.

We thank L.~Lechner, B. Placais, and E. Sonin for fruitful discussions and
S. Iijima, A. Koshio, and M. Yudasaka for the carbon
nanotube material employed in our work. This
work was supported by the Academy of Finland
and by the EU contract FP6-IST-021285-2.


\begin{thebibliography}{99}
\bibitem{NOTE} This upper limit was obtained with the localized contact model and assuming $F=1$ for the contacts.
\end{thebibliography}
\end{document}